\RequirePackage{lineno}
\documentclass[aps,twocolumn,showpacs,preprintnumbers,amsmath]{revtex4}
\usepackage{graphicx}% Include figure files
\usepackage{dcolumn}% Align table columns on decimal point
\usepackage{epstopdf}
\usepackage{color}
\usepackage{bm}% bold math
\definecolor{dgreen}{cmyk}{1.,0.,1.,0.2}        % dark green
\definecolor{orange}{cmyk}{0.,0.353,1.,0.}    % orange

\def\bea {\begin{eqnarray}}
\def\eea {\end{eqnarray}}

\def\be {\begin{equation}}
\def\ee {\end{equation}}

\begin{document}
%\linenumbers
%\begin{frontmatter}
\title{Exploring the initial stage of high multiplicity proton-proton collisions by determining the initial temperature of the Quark-Gluon Plasma }
\author{R. P. Scharenberg{$^1$ }, B. K. Srivastava {$^1$} and C. Pajares{$^2$}}
\medskip
\affiliation{$^1$Department of Physics and Astronomy, Purdue University, West Lafayette, IN-47907, USA\\ 
$^2$Departamento de Fisica de Particulas, Universidale de Santiago de Compostela and Instituto Galego de Fisica de Atlas Enerxias(IGFAE), 15782 Santiago, de Compostela, Spain}

\bigskip

\date{\today}% It is always \today, today
\begin{abstract}
 We have analyzed identified particle transverse momentum spectra in high multiplicity events in $pp$ collisions at LHC energies $\sqrt s $ = 0.9-13 TeV published by the CMS Collaboration using the Color String Percolation Model (CSPM). In CSPM color strings are formed after the collision, which decay into new strings through color neutral $q-\bar{q}$ pairs production. With the increase in the $pp$ collisions energy number of strings grow and randomly statistically overlap producing higher string tension of the composite strings. The net color in the overlap string area is a vector sum of the randomly oriented  strings. The Schwinger color string breaking mechanism produces these color neutral $q-\bar{q}$ pairs at time $\sim $ 1 fm/c, which subsequently hadronize. 

 The initial temperature is extracted both in low and high multiplicity events.The shear viscosity to entropy density ratios $\eta/s$ are obtained as a function of temperature. For the higher multiplicity events at $\sqrt s $ =7 and 13 TeV the initial temperature is above the universal hadronization temperature and is consistent with the creation of deconfined matter. The $\eta/s$ is similar to 
that in Au+Au collisions at $\sqrt {s_{NN}} $ = 200 GeV. The small value of $\eta/s$ above the universal hadronization temperature suggested that the matter is a strongly coupled Quark Gluon Plasma. 
 In these small systems it can be argued that the thermalization is a consequence of the quantum tunneling through the event horizon introduced by the quarks confined in the colliding nucleons and their decelaration due to string formation,
 in analogy to the Hawking-Unruh radiation which provides a stochastic approach to equilibrium. The disk areas cluster on the nucleon transverse collision area. At the $2D$ percolation threshold a macroscopic spanning cluster suddenly occurs at the temperature $T_{i} = T_{h}$, representing a small connected droplet of  $q-\bar{q}$ pairs, the QGP. $T_{h}$ is the universal hadronization temperature $\sim$ 167.7 MeV.
The collision energy dependent buildup of the 2D percolation clusters defines the temperature range $159 \pm 9$ MeV of the crossover transition between hadrons to the QGP in reasonable agreement with the Lattice Quantum Chromo Dynamics ( LQCD) pseudo-critical temperature value of $155 \pm 9$ MeV. 
 
Color String Percolation Model is the new initial stage paradigm for the study of the high density matter produced in $pp$ and $A+A$ collisions. With CSPM we
 can directly explore the thermodynamics of the QGP above the universal hadronization temperature. 
     
\end{abstract}

%\begin{keyword}
 % QGP \sep Deconfinement \sep Shear viscosity \sep Trace anomaly
\pacs{ 25.75.-q, 25.75.Gz, 25.75.Nq, 12.38.Mh} 
%\end{keyword}
\maketitle
%\end{frontmatter}

%\newpage
\section{Introduction}

The observation of high total  multiplicity, high transverse energy, non-jet and  isotropic events led Van Hove to conclude that high energy density events are produced in high energy $\bar{p}p$ collisions \cite{hove}.  In  these  events the transverse energy is proportional to the number of low transverse momentum particles. This basic correspondence has been previously explored over a wide range of the charged particle pseudorapidity density  $ \langle dN_{c}/d\eta \rangle$ in $\bar{p}p$ collisions at center of mass energy $\sqrt {s}$ = 1.8 TeV \cite{E735a}.  The analysis of charged particle transverse momentum in $\bar{p}p$ exhibits flow velocity of mesons and anti-baryons also indicating the possible evidence of QGP formation \cite{muller}. Fermilab experiment E-735 found a multiplicity independent freezout energy density $\sim$ 1.1 $GeV/fm^{3}$ at a temperature of $\sim$ 175 MeV further suggested that deconfined matter is produced in $\bar{p}p$ collisions. The multiplicity dependent freeze out volume was measured using Hanbury-Brown-Twiss pion correlations. Identified particle ratios were used to measure the universal hadronization temperature \cite{E735b}.

The objective of the present work is to explore the initial stage of high energy $pp$ collisions at LHC energies by analyzing the  published CMS data \cite{cms1,cms2} on the transverse momentum spectra of pions using the framework of the clustering of color sources \cite{review} . This phenomenology has been successfully used to describe the initial stages in the soft region in high energy heavy ion collisions \cite{review,nestor,cunq,andres,eos,cpod13,eos2,IS2013,eos3}.  

This requires the measurement of the initial thermalized ( maximum entropy) temperature and the initial energy density at time $\sim 1$ fm/c of the hot matter produced in these high energy nucleon-nucleon collisions. Lattice Quantum Chromo Dynamics simulations (LQCD) indicate that the non-perturbative region of hot QCD matter extends up to temperature of 400 MeV, well above the universal hadronization temperature \cite{latthigh}. 

The initiating colliding quarks and anti-quarks interact to form a large number of color strings. The non-perturbative Schwinger particle creating mechanism in quantum electrodynamics $QED_{2}$, with massless fermions, was derived in an exact gauge invariant calculation \cite{schw}. $QED_{2}$ contains a single space and time coordinate. Confinement, charge screening, asymptotic freedom and the existence of a neutral bound state boson in  $QED_{2}$ closely models $QCD$ . When string color fields are present, the Schwinger $QED_{2}$ string breaking mechanism lifts color neutral $q\bar{q}$ pairs from vacuum \cite{wong}. Schwinger mechanism has also been used in the decay of color flux tubes produced by the quark-gluon plasma for modeling the initial  stages in heavy ion collisions \cite{acta,prc1,prc2}; 

With the increase of the number of strings the strings start to overlap to form clusters on the transverse nucleon interaction plane. At a critical string density, a macroscopic 2D spanning cluster suddenly appears consisting of a connected system of colored disks. When the Schwinger string breaking mechanism produces the  $q\bar{q}$, this connected system represents a small droplet of Quark-Gluon Plasma. With the increasing string density the number of random statistical overlaps of the strings also form composite strings with a higher string tension. The general result is an increase in the average transverse momentum of the  $q\bar{q}$ pairs and a reduction in the multiplicity expected from non-overlapping strings. The buildup of the color cluster structure with increasing beam energies correctly models the crossover phase transition between hadrons and the QGP, in agreement with LQCD. An event horizon is formed by the initiating confined quarks in the colliding nucleons and the nucleon deceleration due to string formation. Barrier penetration of the event horizon leads to a partial loss of information and is the reason for the stochastic thermalization of the $q\bar{q}$ pairs. This maximum entropy temperature $T_{i}$ is increased by the higher string tension of the composite strings.  The combination of the Schwinger $QED_{2}$ string breaking mechanism, the string density dependent cluster formation and the 2D percolation clustering phase transition, are the basic elements of the non-perturbative Color String Percolation Model (CSPM) that has been previously used to study QGP formation in heavy ion ( A+A) collisions \cite{review}. 

The paper is organized as follows. In Sec. II the 2D percolation by random clustering of finite area disks is presented and in Sec. III the SU(3) overlapping of strings are discussed. The measurement of color suppression factor $F(\xi)$ and its relation to temperature are presented in Secs. IV and V. The Hawking-Unruh effect and thermalization is discussed in Sec. VII. Sections VI to VIII deals with the shear viscosity to entropy density ratio, trace anomaly and equation of state respectively.  

\section{2D Percolation by random clustering of finite discs}

\begin{figure}[thbp]
\centering        
\vspace*{-0.2cm}
\includegraphics[width=0.52\textwidth,height=1.5in]{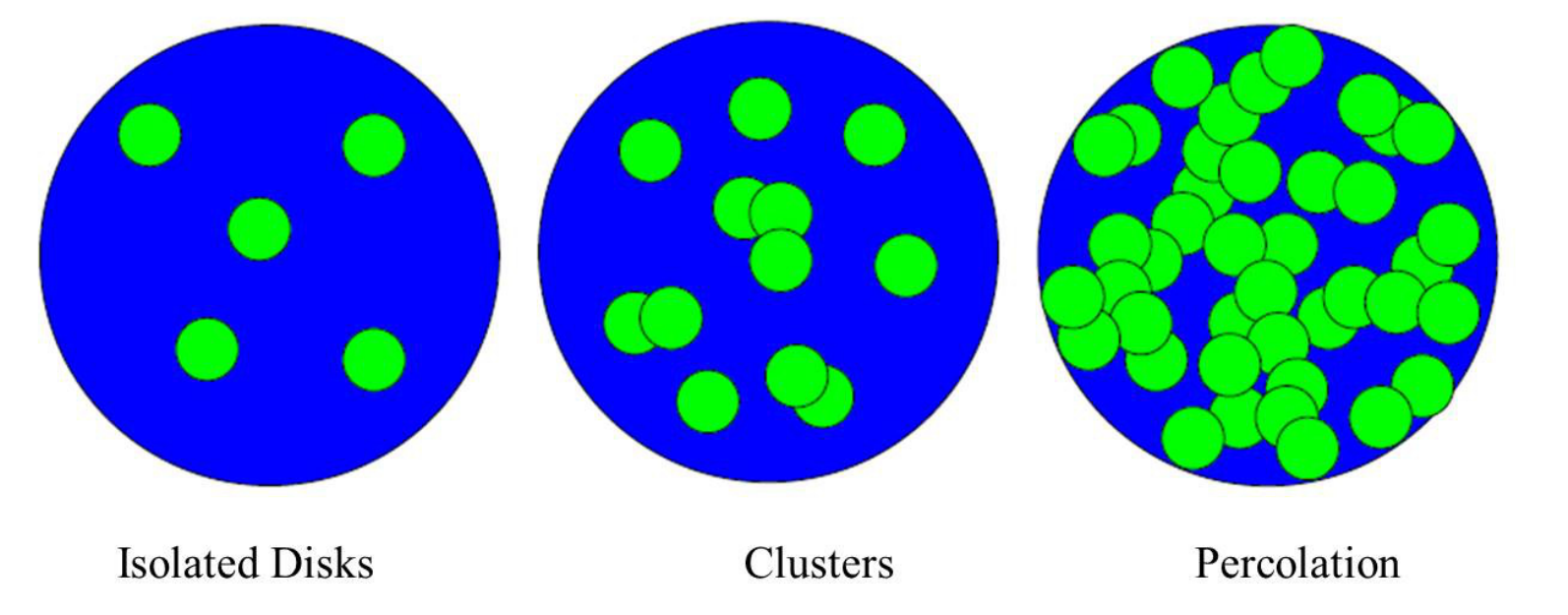}
%\vspace*{-2.0cm}
\caption{Left panel: Disconnected disks, Middle: Cluster formation, Right panel: Overlapping discs forming a cluster of communication \cite{satzbook}.}
\label{cluster}
\end{figure}

A simple example of percolation  is the 2-dimensional continuous percolation \cite{inch}. 
Let us distribute small discs of area $\pi r_{0}^{2}$ randomly on a large surface, allowing overlap between them. As the number of disks increases clusters of overlapping discs start to form. If we regard the disks as small drops of water, how many drops are needed to form a puddle crossing the considered surface? Given $N$ disks, the disk density is $\rho = N/S$ where $S$ is the surface area. The average cluster size increases with $\rho$, and at a certain critical value $\rho_{c}$ the macroscopic cluster spans the whole surface as shown in  Fig.~\ref{cluster} \cite{satzbook}. 
The critical value $\rho_{c}=\frac {1.2}{\pi r_{0}^{2}}$. for the for the onset of 2D continuum percolation was determined by numerical and Monte-Carlo simulations \cite{inch}. In the thermodynamical limit corresponding to  $N \rightarrow \infty $, keeping $\rho$ fixed, the distribution of  overlaps of the disks is Poissonian with a mean value $\xi = \rho \pi r_{0}^2$, $\xi$ being a dimensionless quantity
\begin{equation}
P_{n}= \frac{\xi^{n}}{n!}e^{-\xi}.
\label{per2}
\end{equation}
Hence the fraction of the total area covered by disks is $1-e^{-\xi}$  \cite{inch}. For the critical value of 1.2 $\approx$ 2/3 of the area is covered by discs. This onset of the spanning cluster is used in identifying the connected structure in multi-string excitations. 

\section{The random SU(3) overlapping of single strings and color sources}
Multi-particle production is currently described in terms of color strings stretched between the partons of the projectile and the target. The strings decay into new ones by sea $q-\bar{q}$ production, and subsequently hadronize to  produce the observed hadrons. The color in these strings is confined to small area in the transverse space, $\pi r_{0}^{2}$, with $r_{0} \approx 0.2-0.25$ fm \cite{review}. 

With increasing energy and atomic number of the colliding particles, the number of strings grow and start to overlap, forming clusters, very much like discs in the continuum two dimensional (2D) percolation theory. At a certain critical density $\xi_{c} \sim$ 1.2 a macroscopic cluster appears, which marks the percolation phase transition. For nuclear collisions, this density corresponds to $\xi = N_{s} \frac {S_{1}}{S_{A}}$ where $N_{s}$ is the total number of strings created in the collision, each one of area $S_{1} = \pi r_{0}^{2}$ and $S_{A}$ corresponds to the nuclear overlap area. 

The percolation theory governs the geometrical pattern of string clustering. Its observable implications, however, require the introduction of some dynamics in order to describe the behavior of the cluster formed by several overlapping strings. We assume that a cluster of $n$ strings behaves as a single string with an energy-momentum that corresponds to the sum of the energy-momenta of the overlapping strings and with a higher color field, corresponding to the vectorial sum of color charges of each individual string  $\vec{Q_{1}}$. The resulting color field covers the area $S_{n}$ of the cluster. As $\vec{Q_{n}^{2}} = (\sum_{1}^{n}\vec{Q_{1}})^{2}$, and the individual string colors may be oriented in an arbitrary manner respective to each other, the average $\vec{Q_{1i}}\vec{Q_{1j}}$ is zero, and $\vec{Q_{n}^2} = n \vec{Q_{1}^2} $. $\vec{Q_{1}}$ depends also on the area $S_{1}$ of each individual string that comes into the cluster, as well as on the total area of the cluster $S_{n}$ \cite{pajares1,pajares2}
\begin{equation}
Q_{n} = \sqrt {\frac {n S_{n}}{S_{1}}}Q_{1}.
\end{equation} 

We take $S_{1}$ constant and equal to a disk of radius $r_{0}$. $S_{n}$ corresponds to the total area occupied by $n$ disks, which of course can be different for different configurations even if the clusters have the same number of strings.
If the strings are just touching each other, $S_{n} = n S_{1}$ and $Q_{n}=nQ_{1}$,
so the strings act independently to each other. On the contrary, if they fully overlap $S_{n} = S_{1}$ and $Q_{n}= \sqrt{n} Q_{1}$, then we obtain a reduction in the color charge. Knowing the color charge $\vec{Q_{n}}$, one can compute the multiplicity $\mu_{n}$ and the mean transverse momentum squared $\langle p_{t}^{2} \rangle_{n}$ of the particles produced by a cluster, which are proportional to the color charge and color field, respectively,
\begin{equation}
\mu_{n} = \sqrt {\frac {n S_{n}}{S_{1}}}\mu_{1};\hspace{5mm}
%\end{equation}
%\begin{linenomath*}
 %\begin{equation}                                      
\langle p_{t}^{2} \rangle_{n} = \sqrt {\frac {n S_{1}}{S_{n}}} {\langle p_{t}^{2} \rangle_{1}},
\label{mu}
\end{equation} 
where $\mu_{1}$ and $\langle p_{t}^{2}\rangle_{1}$ are the mean multiplicity and transverse momentum squared of particles produced from a single string. In the
thermodynamic limit Eq.~(\ref{mu}), can be written as \cite{pajares1,pajares2}

\begin{equation}
  \mu_{n} \langle p_{t}^{2}\rangle_{n} = n\mu_{1}\langle p_{t}^{2} \rangle_{1};\hspace{5mm}
  \frac {\mu_{n}}{\langle p_{t}^{2}\rangle_{n}} = \frac{S_{n}}{S_{1}} \frac {\mu_{n}}{\langle p_{t}^{2} \rangle_{1}}.
  \label{mu2}
 \end{equation} 

The first relation denotes that the product is an extensive quantity, while the second one indicates that each cluster satisfies a scaling law that is nothing but Gauss theorem.  
%From the Schwinger formula , one obtains $\mu_{1}= S_{1} \langle p_{t}^{2} \rangle_{1}$. 
Moreover, in the limit of high density $\xi$ , one obtains the average value for $nS_{1}/S_{n}$  
\begin{equation}
\left \langle{n\frac {S_{1}}{S_n}}\right \rangle  =  {\frac {\xi}{1-e^{-\xi}}}\equiv\frac {1}{F(\xi)^{2}}.
  \label{xi2}
 \end{equation} 
and the Eqs.~(\ref{mu2}) and (\ref{xi2})  transform into the analytical ones 

\begin{equation}
  \mu_{n}= n F(\xi)\mu_{1};\hspace{5mm}
  \langle p_{t}^{2}\rangle_{n} ={\langle p_{t}^{2} \rangle_{1}}/F(\xi).
  \label{mu3}
 \end{equation} 
 \section{The measurement of the color suppression factor  $F(\xi)$}

In order to compute the transverse momentum distribution, we make use of the  parameterization of the experimental data of $p_{t}$ distribution in low energy ${\it pp}$ collisions  $\sqrt s$ = 200 GeV  

\begin{equation}
  d^{2}N_{c}/dp_{t}^{2} = a/(p_{0}+p_{t})^{\alpha},
  \label{spectra1}
\end{equation}
where $a$ is the normalization factor, $p_{0}$ and $\alpha$ are fitting parameters with $p_{0}$= 1.98 and  $\alpha$ = 12.87 \cite{eos}. This parameterization is used in high multiplicity ${\it pp}$ collisions to take into account the interactions of the strings \cite{review}.

\begin{equation}
p_{0} \rightarrow  p_{0} \left(\frac {\langle nS_{1}/S_{n} \rangle_{pp}^{mult}}{\langle n S_{1}/S_{n} \rangle_{pp}}\right)^{1/4},
\end{equation}
In $pp$ collision at low energies only two strings are exchanged with small probability of interactions, so that $\langle n S_{1}/S_{n}\rangle_{pp } \simeq$1.
which transforms Eq.~(\ref{spectra1} ) into
\begin{equation} 
  \frac{d^{2}N_{c}}{dp_{T}^{2}} = \frac{a}{(p_{0} \sqrt {F(\xi)_{pp}/F(\xi)_{pp}^{mult}}+{p_{T}})^{\alpha}},
  \label{spectra2} 
\end{equation}
where $F(\xi)_{pp}^{mult}$ is the multiplicity dependent color suppression factor.  In $pp$ collisions $F(\xi)_{pp} \sim$ 1 at low energies due to the low overlap probability. The spectra were fitted using  Eq.~(\ref{spectra2}) in the softer sector with $p_{t}$ in the range 0.12-1.0 GeV/c.  

In the thermodynamic limit the color suppression factor $F(\xi)$ is related to the percolation density parameter $\xi $ given by 
\begin{equation} 
F(\xi) = \sqrt {\frac {1-e^{-\xi}}{\xi}}
\end{equation}

\begin{figure}[thbp]
\centering        
\vspace*{-0.2cm}
\includegraphics[width=0.50\textwidth,height=3.0in]{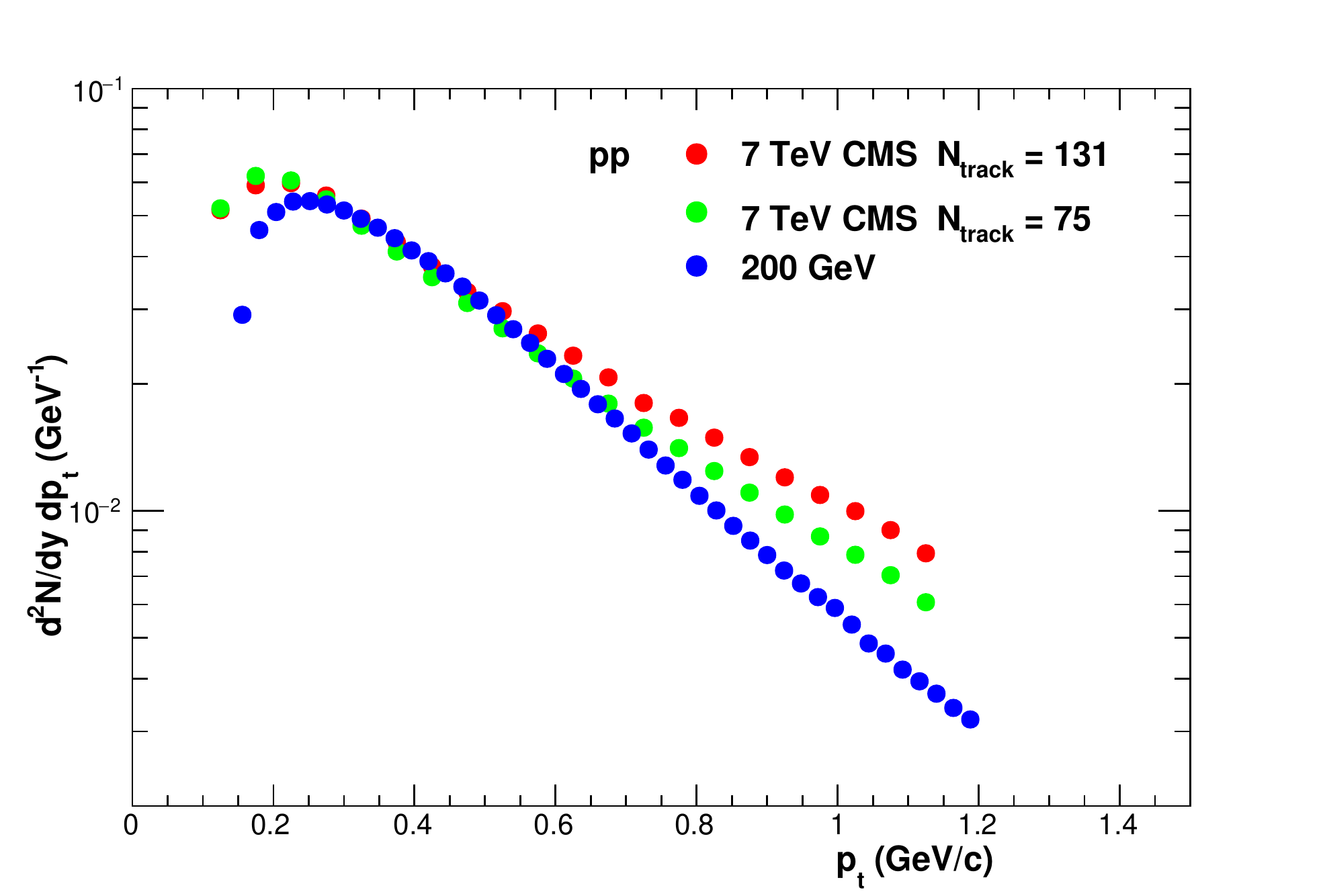}
\vspace*{-0.8cm}
\caption{ Transverse momentum spectra of $pions$ from CMS experiment at $\sqrt s $= 7 TeV for two different multiplicity cuts $N_{track} = 131 $ (red solid circle) and $N_{track}$ = 75 (green solid circle) \cite{cms1}. For comparison purpose the $p_{t}$ spectra from $pp$ at $\sqrt s $ = 200 GeV is also shown (solid blue circle) \cite{review}.
}
\label{ptspectra}
\end{figure}

%\section{Results and discussions}
In the present work we have extracted $F(\xi)$ in high multiplicity events in $pp$ collisions using CMS data from the transverse momentum spectra of pions at $\sqrt s$ =  0.9, 2.76, 7 and 13 TeV \cite{cms1,cms2}. Figure ~\ref{ptspectra} shows a transverse momentum spectra for two multiplicity cuts at $\sqrt s$ = 7 TeV. For comparison purpose the spectra from  $pp$ collisions $\sqrt s$ = 200 GeV is also shown. The spectra becomes harder for higher multiplicity cuts. This is due to the fact that high string density color sources are created in the higher multiplicity events. 
\begin{figure}[thbp]
\centering        
\vspace*{-0.2cm}
\includegraphics[width=0.50\textwidth,height=3.0in]{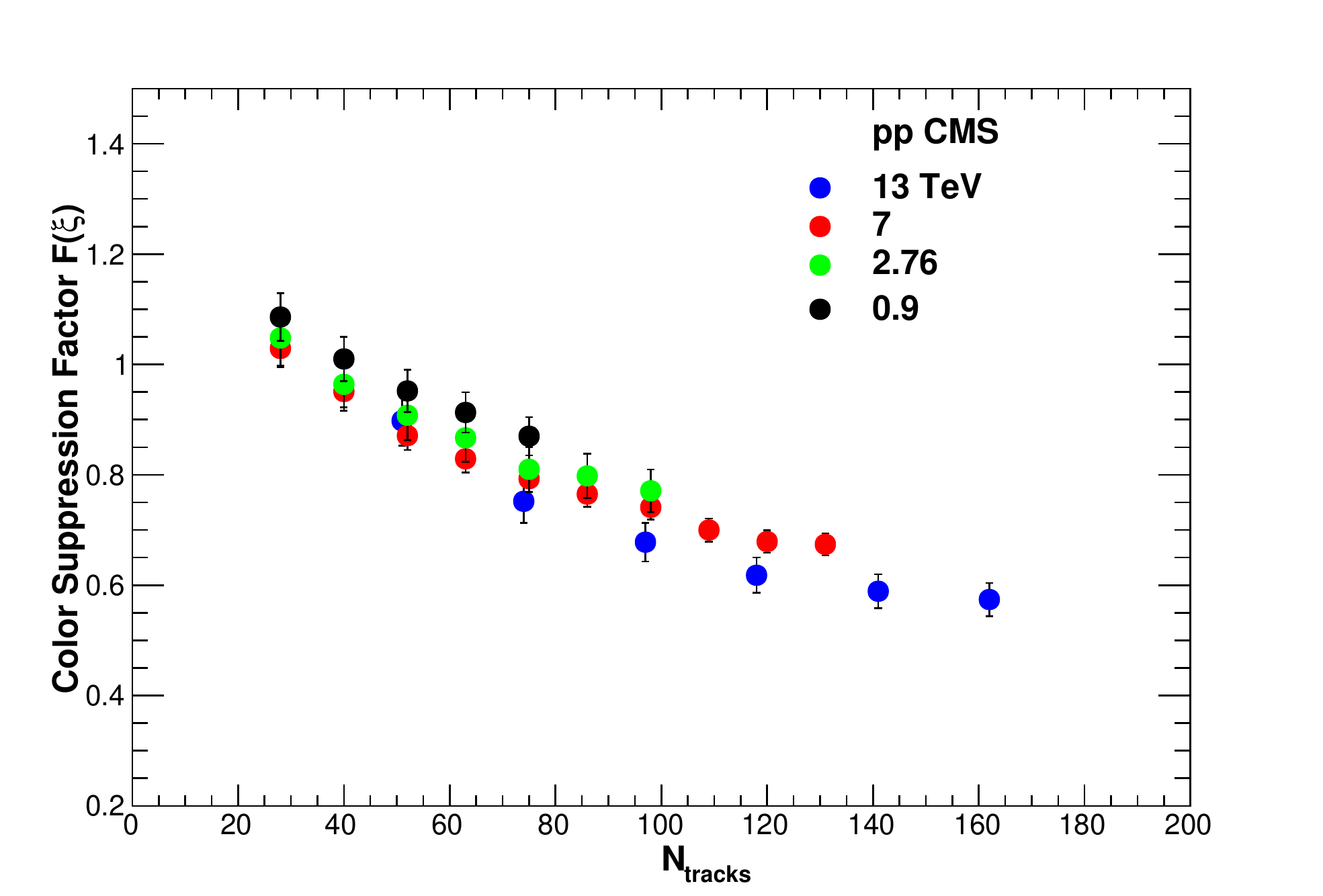}
\vspace*{-0.5cm}
\caption{Color Suppression Factor $F(\xi)$ in ${\it pp}$ collisions vs $N_{tracks}$}
\label{fxipp2_1}
\end{figure}  

Figure~\ref{fxipp2_1} shows the extracted value of $F(\xi)$ as a function of $N_{tracks}$ from CMS experiment for $\sqrt {s}$ = 0.9 - 13 TeV. It is observed that the $F(\xi)$ has lower value for high multiplicity events and increases for low multiplicity events. To compare with the heavy ions results with the $pp$ one need to normalized $N_{tracks}$  with the interaction area in $pp$ collisions. 
\begin{figure}[thbp]
\centering        
\vspace*{-0.2cm}
\includegraphics[width=0.50\textwidth,height=3.0in]{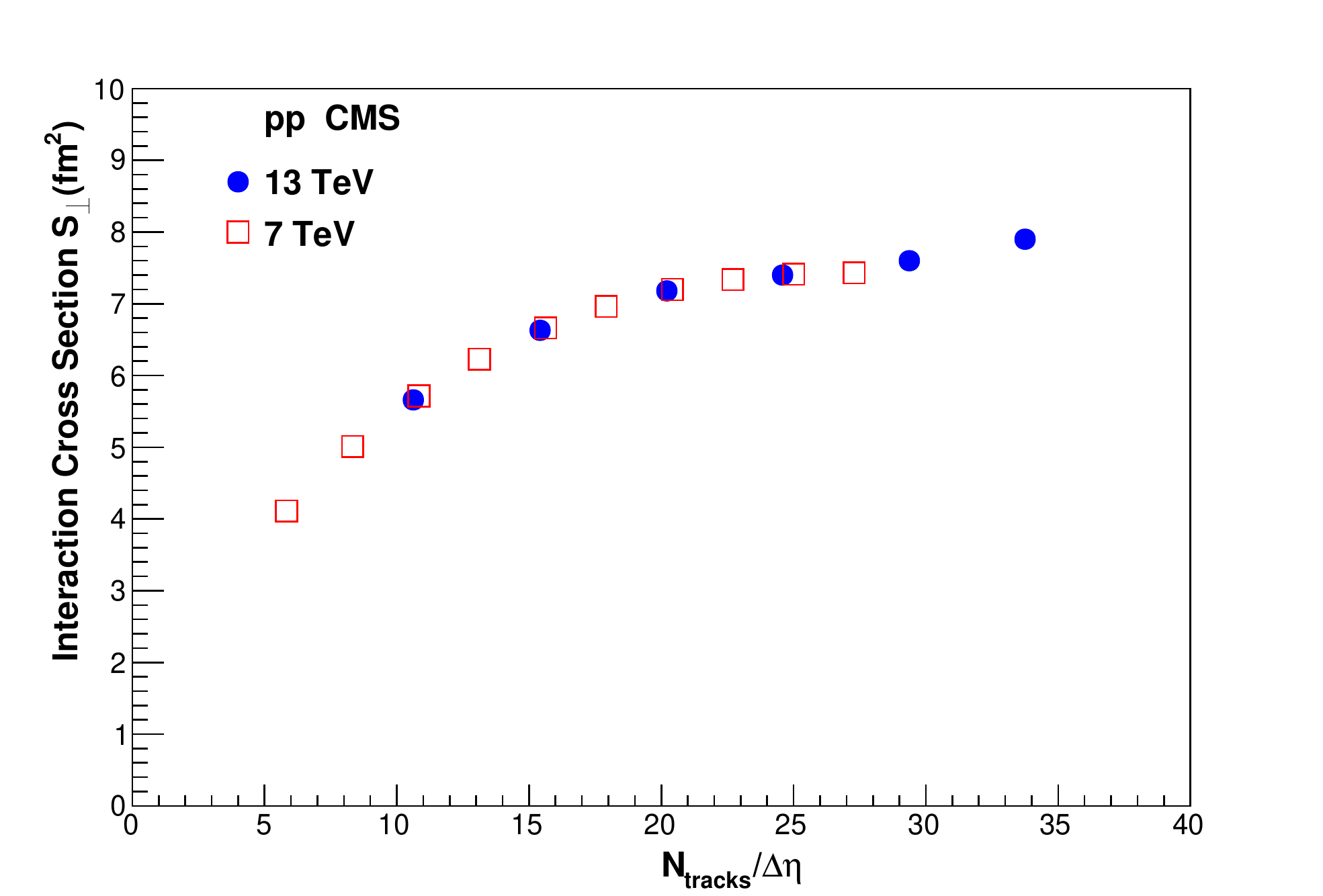}
\vspace*{-0.5cm}
\caption{Interaction cross section $S_{\perp}$ vs $N_{tracks}/\Delta\eta$. $S_{\perp}$ is obtained using IP-Glasma model \cite{cross}.}
\label{cross}
\end{figure} 
The interaction area $S_{\perp} = \pi R^{2}_{pp}$   has been computed  in the IP-Glasma model, where $R_{pp}$ is the interaction radius \cite{cross} . This is based on an impact parameter description of $ {\it pp} $ collisions, combined with an underlying description of particle production based on the theory of Color Glass Condensate \cite{cross}.  For the higher multiplicity events the interaction radius $R_{pp}$ is approximately a linear function of the charged particle multiplicity. In the IP-Glasma model $R_{pp}$ is dependent on gluon multiplicity \cite{cross}

\begin{equation}
  R_{pp} = f_{pp}(dN_{g}/dy)^{1/3}
\label{cross2}
\end{equation}
\[
 f_{pp} = \left\{ \begin{array}{ll}
                      0.387+0.0335x+0.274x^{2}-0.0542x^{3},   &   {\rm if} x < 3.4,\\
                       1.538                                 &   {\rm if} x \geq 3.4. 
                     \end{array}          
  \right.
\]
The gluon multiplicity $dN_{g}/dy$ is related to the number of tracks seen in the CMS experiment by
\begin{equation}
dN_{g}/dy \approx (3/2) \frac {1}{\Delta\eta} N_{track},
\label{cross3}
\end{equation}
where $\Delta\eta \sim $ 4.8 units of pseudorapidity.  The interaction cross section $S_{\perp}$  as a function of $N_{tracks}/\Delta\eta$, using Eqs.~(\ref{cross2}), is shown in Fig.~\ref{cross}. $S_{\perp}$ increases with the multiplicity and for very high multiplicities it is approximately constant.

\begin{figure}[thbp]
\centering        
\vspace*{-0.2cm}
\includegraphics[width=0.50\textwidth,height=4.0in]{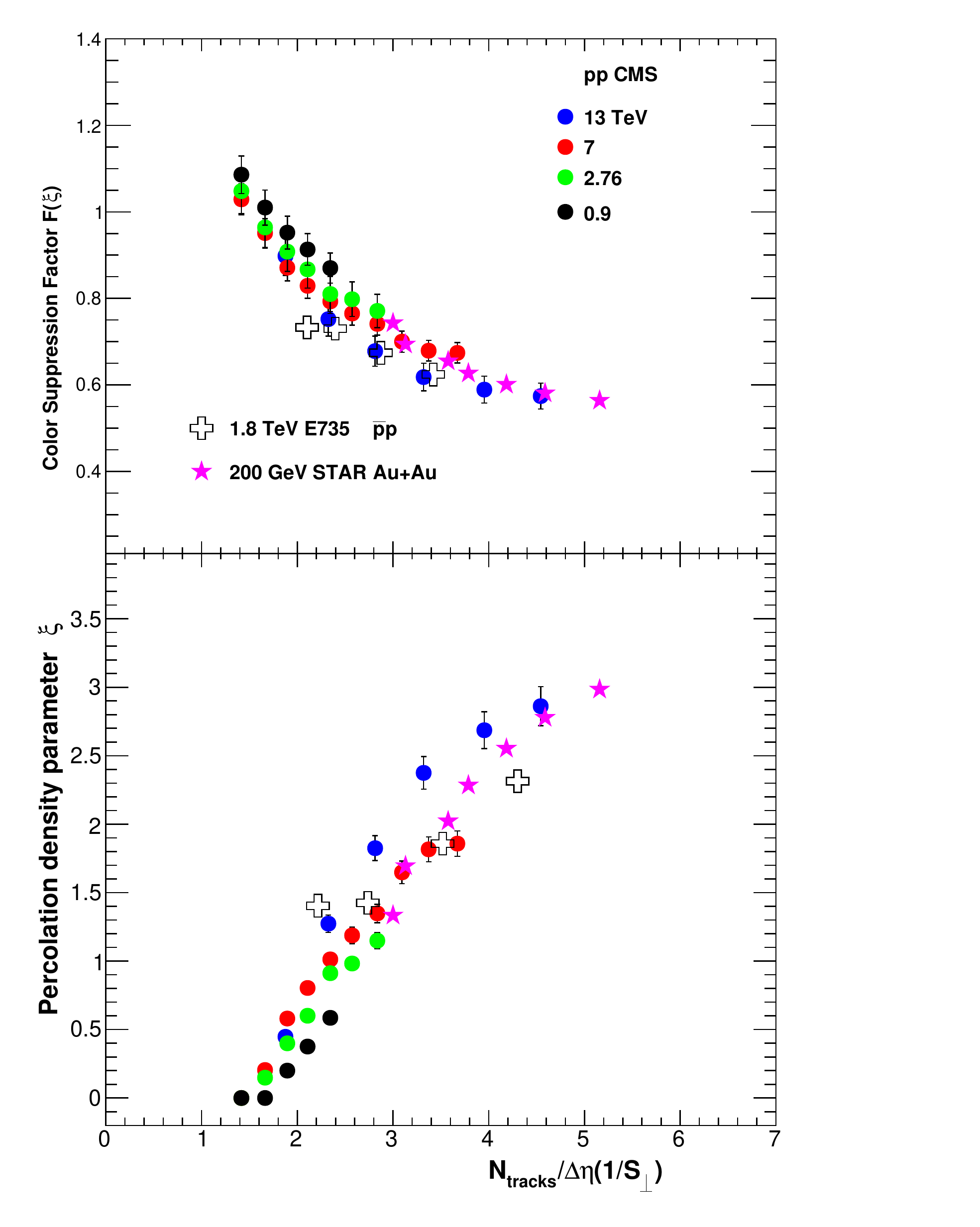}
\vspace*{-0.5cm}
\caption{
(a). Color Suppression Factor $F(\xi)$ in ${\it pp}$, ${ \bar p p}$ and $Au+Au$ collisions vs $N_{tracks}/\Delta\eta$ scaled by the transverse area $S_{\perp}$. For ${\it pp}$ and ${ \bar p p}$ collisions $S_{\perp}$ is multiplicity dependent as obtained from IP-Glasma model \cite{cross}. In case of $Au+Au$ collisions the nuclear overlap area was obtained using Glauber model \cite{glauber}.\newline 
(b). Percolation density parameter $\xi$ as a function of $N_{tracks}/\Delta\eta$ .
}
\label{fxipp2}
\end{figure}

Figure~\ref{fxipp2} shows the extracted value of $F(\xi)$ as a function of $N_{tracks}/\Delta\eta$ scaled by the interaction area $S_{\perp}$ from the CMS experiment for $\sqrt {s}$ = 0.9 - 13 TeV. 
$N_{tracks}$ is the total charged particle multiplicity in the region $|\eta| < 2.4$ with $\Delta\eta \sim$ 4.8 units of pseudorapidity.

The results from FNAL (Fermi National Accelerator Laboratory) E735 experiment on $\bar{p}p$ collisions at $\sqrt {s}$ = 1.8 TeV is also shown in Fig.~\ref{fxipp2} \cite{e735}. In the E735 experiment the total charged particle multiplicity was 10 $ < N_{c} < $ 200 in the pseudorapidity range $|\eta| < $3.25 with $\Delta\eta \sim$ 6.5 units of pseudorapidity. It is observed that the E735 results follow the trend as seen in CMS data. The decrease in $F(\xi)$ for high multiplicity events is due to the high string density created in theses events. 
 \section{Relationship of  $F(\xi)$ to the initial temperature} 

The connection between $F(\xi)$ and the temperature $T(\xi)$ involves the Schwinger mechanism for particle production \cite{review, schw, pajares3}.
The Schwinger distribution for massless particles is expressed in terms of $p_{t}^{2}$ \cite{review, pajares3}
\begin{equation}
  dn/d{p_{t}^{2}} \sim exp(-\pi p_{t}^{2}/x^{2})
  \label{qed}
\end{equation}
where the average value of the string tension is  $\langle x^{2} \rangle$. 
The tension of the macroscopic cluster fluctuates around its mean value because the chromo-electric field is not constant.
The origin of the string fluctuation is related to the stochastic picture of 
the QCD vacuum. Since the average value of the color field strength must 
vanish, it cannot be constant but changes randomly from point to point~\cite{bialas}. Such fluctuations lead to a Gaussian distribution of the string tension
\begin{equation}
\frac{dn}{dp_{t}^{2}} \sim \sqrt \frac{2}{<x^{2}>}\int_{0}^{\infty}dx exp\left(-\frac{x^{2}}{2<x^{2}>}\right) exp(-\pi \frac{p_{t}^{2}}{x^{2}})
\end{equation}
 which gives rise to  thermal distribution~\cite{bialas}
\begin{equation}
\frac{dn}{dp_{t}^{2}} \sim exp\left(-p_{t} \sqrt {\frac {2\pi}{\langle x^{2} \rangle}}\right),
\label{bia}
\end{equation}
with $\langle x^{2} \rangle$ = $\pi \langle p_{t}^{2} \rangle_{1}/F(\xi)$. 
The temperature is expressed as~\cite{pajares3,eos}  
\begin{equation}
T(\xi) =  {\sqrt {\frac {\langle p_{t}^{2}\rangle_{1}}{ 2 F(\xi)}}}.
\label{temp}
\end{equation} 
We will adopt the point of view that the universal hadronization
%experimentally determined chemical freeze-out
 temperature is a good measure of the upper end of the cross over phase transition temperature $T_{h}$ \cite{bec1}. 
%$T_{c}$~\cite{braun}. 
The single string average transverse momentum  ${\langle p_{t}^{2}\rangle_{1}}$ is calculated at $\xi_{c}$ = 1.2 with the  universal hadronization temperature 
$T_{h}$ 167.7 $\pm$ 2.6 MeV~\cite{bec1}. This gives \mbox{$ \sqrt {\langle {p_{t}^{2}} \rangle _{1}}$  =  207.2 $\pm$ 3.3 MeV}. 

In this way at \mbox{$\xi_{c}$ = 1.2} the connectivity percolation transition at $T(\xi_{c})$ models the thermal deconfinement transition. The~temperature obtained using Equation (\ref{temp}) was $\sim$193.6 MeV for Au+Au collisions at \mbox{$\sqrt{s_{NN}}$ = 200 GeV} in reasonable agreement with $T_{i}$ = 221 $\pm$ $19^{stat} \pm 19^{sys}$ MeV from the enhanced direct photon experiment measured by the PHENIX Collaboration~\cite{phenix}.  
For Pb-Pb collisions at \mbox{$\sqrt{s_{NN}}$ = 2.76 TeV} the temperature is 
$\sim$262.2 MeV for 0--5$\%$ centrality ~\cite{eos}. 
The direct photon measurements from ALICE gives the temperature of $T_{i}$  = 304 $\pm 51$ MeV \cite{alicetemp}.
The agreement with the measured temperature shows that the temperature obtained using Eq. (\ref{temp}) can be termed as the initial temperature of the percolation cluster.
\begin{figure}[thbp]
\centering        
\vspace*{1.0cm}
\includegraphics[width=0.50\textwidth,height=3.0in]{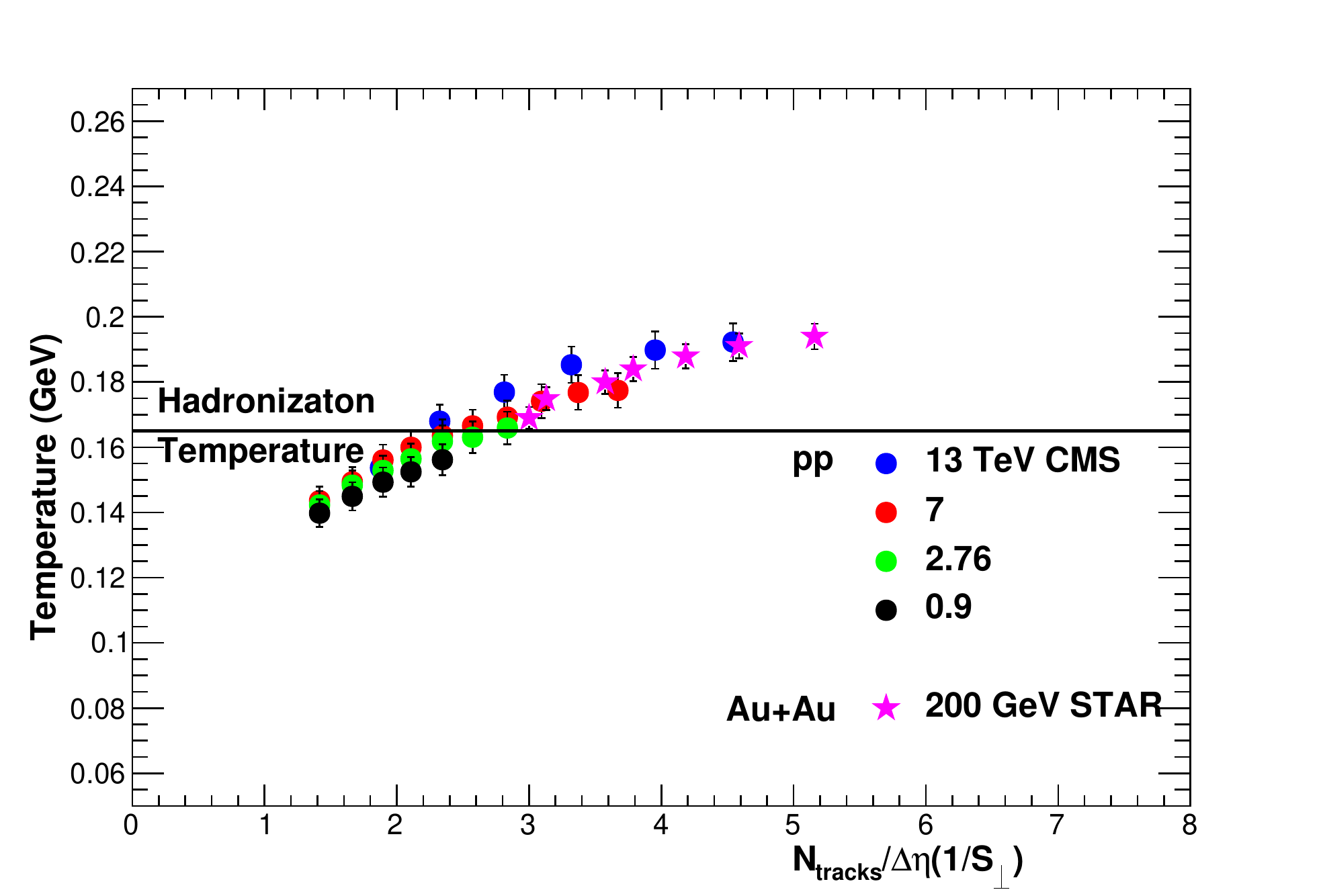}
\vspace*{-0.5cm}
\caption{ Temperature vs $N_{tracks}/\Delta\eta$ scaled by  $S_{\perp}$ from ${\it pp}$ and $Au+Au$ collisions. The horizontal  line  $\sim 165 $ MeV is the universal hadronization temperature \cite{bec1}. } 
\label{tempfig}
\end{figure}
% Unruh
Figure~\ref{tempfig} shows a plot of temperature as a function of $N_{tracks}/\Delta\eta$ scaled by $S_{\perp}$. Temperature from both hadron-hadron and nucleus-nucleus  collisions fall on a universal curve when multiplicity is scaled by the transverse interaction area. The horizontal line at $\sim $ 167.7 MeV is the universal hadronization temperature obtained from the systematic comparison of the statistical thermal model parametrization of hadron abundances measured in high energy  $e^{+}e^{-}$, $pp$ and A+A collisions \cite{bec1}. In Fig.~\ref{tempfig} for $\sqrt {s}$ = 7 and 13 TeV higher multiplicity cuts show temperatures above the hadronization temperature and similar to those observed in Au+Au collisions at $\sqrt {s_{NN}}$ = 200 GeV.
\section{Hawking-Unruh effect, event horizon and  the stochasitc thermalization}

Recently, it has been suggested that fast thermalization in heavy ion collisions can occur through the existence of an event horizon caused by a rapid deceleration of the colliding nuclei \cite{casto1,alex}. The thermalization in this case is due to the Hawking-Unruh effect \cite{hawk,unru}. 

 It is well known that the black holes evaporates by quantum pair production and behave as if they have an effective temperature of 
\begin{equation}
T_{H}= \frac {1}{8\pi GM},
\end{equation}
where 1/4GM is the acceleration of gravity at the surface of a black hole of mass M. The rate of pair production in the gravitational background of the black hole can be evaluated by considering the tunneling through the event horizon. Unruh showed that a similar effect arises in a uniformly accelerated frame, where an observer detects the thermal radiation with the  temperature T =a/2,
where $a$ is the acceleration. Similarly, in hadronic interactions the probability to produce states of masses M due to the chromoelectric field E and color charge is given by the Schwinger mechanism \cite{schw}
\begin{equation}
W_{M} \sim \exp (\frac {-\pi M^{2}}{gE})\sim \exp(-M/T),
\end{equation}
which is similar to the Boltzmann weight in a heat bath with an effective temperature
\begin{equation}
T = \frac {a}{2\pi}, a= \frac {2gE}{M}.
\end{equation}
In CSPM the strong color field inside the large cluster produces de-acceleration of the primary $q \bar q$ pair which can be seen as a thermal temperature by means of the Hawking-Unruh effect \cite{hawk,unru}. This implies that the radiation temperature is determined by the transverse extension of the color flux tube/cluster in terms of the string tension \cite{casto1,alex}. 
\begin{equation}
T =  {\sqrt \frac {\sigma}{2\pi}}
\label{haw}
\end{equation}
The string percolation density parameter $\xi$ which characterizes the percolation clusters measures the initial temperature via the color reduction factor F($\xi$). Since the cluster covers most of the interaction area, this temperature becomes a global temperature.

\section{The temperature dependence of shear viscosity to entropy density ratio}
The relativistic kinetic theory relation for the shear viscosity over entropy density ratio, $\eta/s$ is given by~\cite{gul1, gul2}
\begin{equation}
\frac {\eta}{s} \simeq \frac {T \lambda_{mfp}}{5}   
\label{gulvis}  
\end{equation}
where \emph{T} is the temperature and $\lambda_{mfp}$ is the mean free path given by
\begin{equation}
\lambda_{mfp} \sim \frac {1}{(n\sigma_{tr})}
\label{gulvis2}  
\end{equation}
$\it n $ is the number density of an ideal gas of quarks and gluons and $\sigma_{tr}$ the transport cross section for these constituents. 
In CSPM the number density is given by the effective number of sources per unit volume 
\begin{equation}
n = \frac {N_{sources}}{S_{N}L},
\label{source}
\end{equation}
where \emph{L} is the longitudinal extension of the source \emph{L} $\simeq$ 
1.0 $\it  fm $~\cite{,wong,pajares3}. The area occupied by the strings is related to the percolation density parameter $\xi$ through the relation $(1-e^{-\xi})S_{N}$. Thus the effective no. of sources is given by the total area occupied by the strings divided by the effective area of the string $S_{1}F(\xi) $. 
\begin{equation}
N_{sources} = \frac {(1-e^{-\xi}) S_{N}}{S_{1} F(\xi)}. 
\label{vie2}
\end{equation}
In general $N_{sources}$ is  smaller than the number of single strings. $N_{sources}$ equals the number of strings $N_{s}$ in the limit of $\xi $ = 0. 
The number density of sources from Eqs. (\ref{vie2}) and (\ref{source}) becomes
\begin{equation}
n = \frac {(1-e^{-\xi})}{S_{1}F(\xi) L}.
\end{equation}
The transport cross section $\sigma_{tr}$ is the transverse area of the effective string $S_{1}F(\xi)$. Thus $\sigma_{tr}$ is directly proportional to $\frac {1}{T^{2}}$, which is in agreement with the estimated dependence of $\sigma_{tr}$ on the temperature \cite{gul2}. The mean free path is given by
\begin{equation}
\lambda_{mfp} = {\frac {L}{(1-e^{-\xi})}}. 
\end{equation}
For large values of $\xi$ the $\lambda_{mfp}$ reaches a constant value.
$\eta/s$ is obtained from $\xi$ and the temperature
\begin{equation}
\frac {\eta}{s} ={\frac {TL}{5(1-e^{-\xi})}} 
\label{viseq}
\end{equation}
 Below $\xi_{c}$ , as the temperature increases, the string density increases and the area is filled rapidly  and $\lambda_{mfp}$ and $\eta/s$ decrease sharply. Above $\xi_{c}$, more than 2/3 of the area are already covered by strings, and therefore the area is not filling as fast and the relatively small decrease of $\lambda_{mfp}$ is compensated by the rising of temperature, resulting in a smooth increase of $\eta/s$. The behavior of $\eta/s$ is dominated by the fractional area covered by strings. This is not surprising because $\eta/s$ is the ability to transport momenta at large distances and that has to do with the density of voids in the matter.
 
\begin{figure}[thbp]
%\centering        
\vspace*{-0.2cm}
\includegraphics[width=0.60\textwidth,height=5.0in]{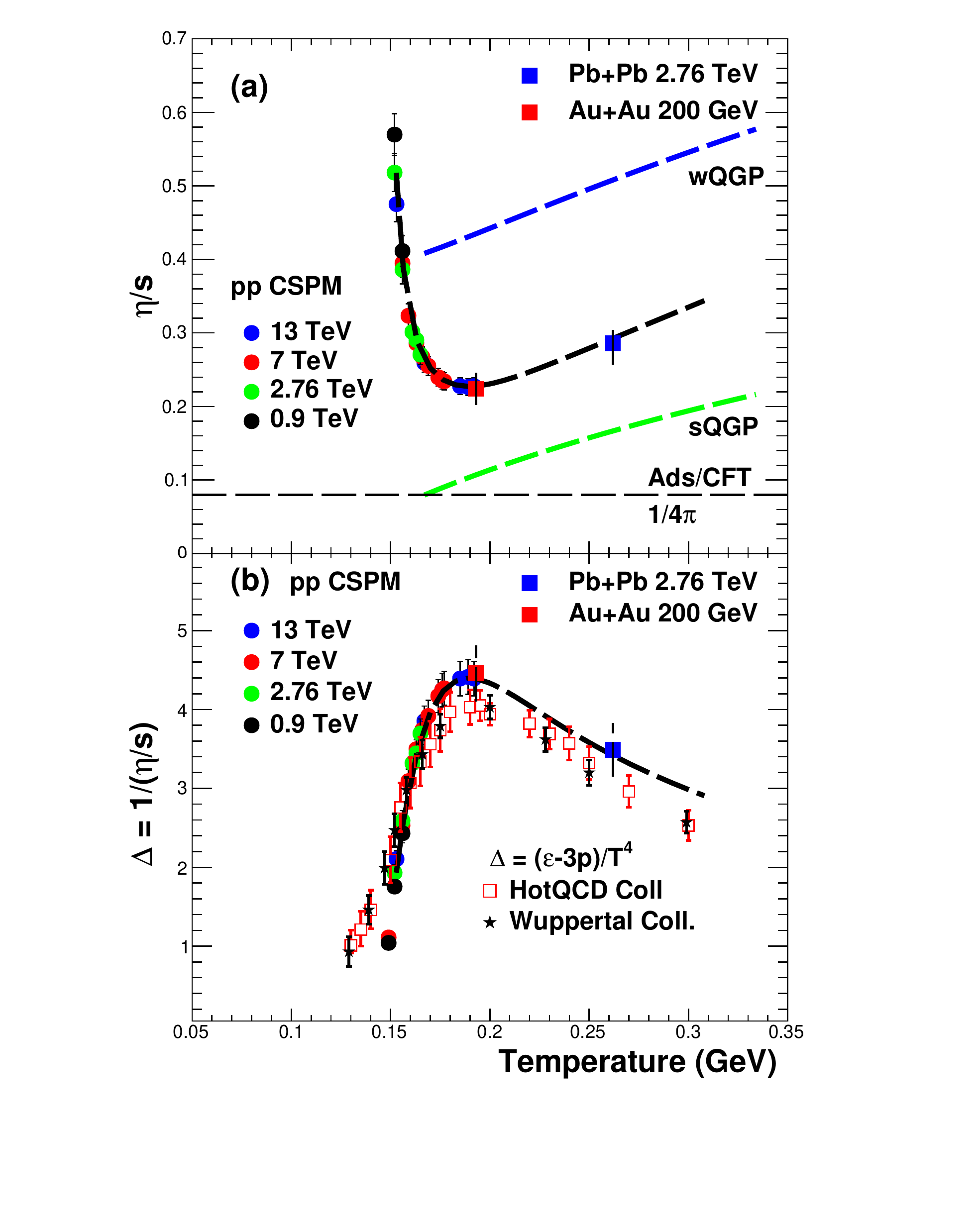}
\vspace*{-2.0cm}
\caption{ (a). $\eta/s$ as a function of temperature T using Eq. (\ref{viseq}) for $\sqrt s $ = 0.9, 2.76, 7 and 13 TeV. The lower bound shown is given by the AdS/CFT \cite{kss}. For comparison purposes the results from Au+Au and Pb+Pb at $\sqrt s_{NN} $ = 200 GeV and 2.76 TeV respectively are also shown as solid squares for 0-5 \% centrality \cite{review}. \newline
 (b). The trace anomaly $\Delta =(\varepsilon-3p)/T^{4}$ vs temperature. Blue open squares are from HotQCD Collaboration \cite{lattice12}. Black stars are from Wuppertal Collaboration \cite{wuppe}. The CSPM results are obtained as $\Delta = 1/(\eta/s)$ \cite{review}. The black dashed line both in (a) and (b) corresponds to extrapolation from CSPM at higher temperatures.} 
\label{viscosity}
\end{figure}

Figure~\ref{viscosity}(a) shows $\eta/s$ as a function of the temperature \cite{eos2}. The lower bound shown in Fig.~\ref{viscosity}(a) is given by the AdS/CFT conjecture \cite{kss}.  The results from $pp$ collisions from $\sqrt {s}$ = 13 TeV shows a very small  $\eta/s$ and that is 2.7 times the AdS/CFT conjectured lower bound $1/4\pi$. The theoretical estimates of $\eta/s$ has been obtained as a function of temperature for both the weakly interacting (wQGP) and strongly interacting (sQGP) coupled QCD plasma are shown in Figure~\ref{viscosity}(a)~\cite{gul1}. 

$\eta/s$ has also been obtained in several other calculations for pure glue matter~\cite{toneev}, in the semi quark gluon plasma~\cite{hidaka} and in a quasi particle description~\cite{bluhm}. In pure SU(3)  gluodynamics a conservative upper bound for $\eta/s$ was obtained  $\eta/s$ = 0.134(33) at $T=1.65 T_{c}$~\cite{meyer}.  In the quasi particle approach also low $\eta/s$$\sim$0.2 is obtained for T $ > 1.05 T_{c}$ and rises very slowly with the increase in temperature~\cite{peshier}. In CSPM also $\eta/s$ grows slowly with temperature as 0.16$T$/$T_{c}$.

\section{The temperature dependence of the trace Anomaly}

The trace anomaly is the expectation value of the trace of the energy-momentum tensor, $\langle \Theta_{\mu}^{\mu}\rangle = (\varepsilon-3p)$, which measures the deviation from conformal behavior and thus identifies the interaction still present in the medium \cite{cheng}. The inverse of $\eta/s$ also measures how strong are the interactions in the medium and therefore we expect a similar behavior as seen in the trace anomaly \cite{eos3}. Figure~\ref{viscosity}(b) shows 1/($\eta/s$) and the dimensionless quantity, $(\varepsilon-3p)/T^{4}$, obtained from lattice simulations \cite{lattice12}.
We make the ${\it ansatz}$ that the temperature dependence of inverse of $\eta/s$ is equal to the dimensionless trace anomaly $\Delta = (\varepsilon-3p)/T^{4}$ \cite{eos3}. The inverse of $\eta/s$ is in qualitative agreement with $\Delta$ over a wide range of temperatures with the LQCD simulations\cite{lattice12}.
The maximum in $\Delta$ corresponds to the minimum in $\eta/s$. Both $\Delta$ and $\eta/s$ describe the transition from a strongly coupled QGP to a weakly coupled QGP. This result is shown in Fig.~\ref{viscosity}(b).

We are not aware of any theoretical work which directly relates the trace anomaly with the shear viscosity to entropy density ratio. However, the bulk viscosity $\zeta$ is related to both $\Delta$ and $\eta$~\cite{teany}. A~detailed study based on low energy theorems and the lattice result for $\Delta$ shows that $\zeta/s$ rises very fast close to the critical temperature in such a way that its value at temperatures higher than $T >1.1  T_{c}$ is quite negligible~\cite{karsch}. It was observed that $\zeta$ scales as $\alpha^{4}_{s} \eta$ where $\alpha_{s}$ is the coupling constant. The~trace anomaly $\Delta$ is proportional to $\alpha^{2}_{s}$~\cite{teany}. 

\section {The temperature dependence of the Equation of State $C_{s}^{2}$} 
\begin{figure}[thbp]
\centering        
\vspace*{-0.2cm}

\includegraphics[width=0.55\textwidth,height=3.0in]{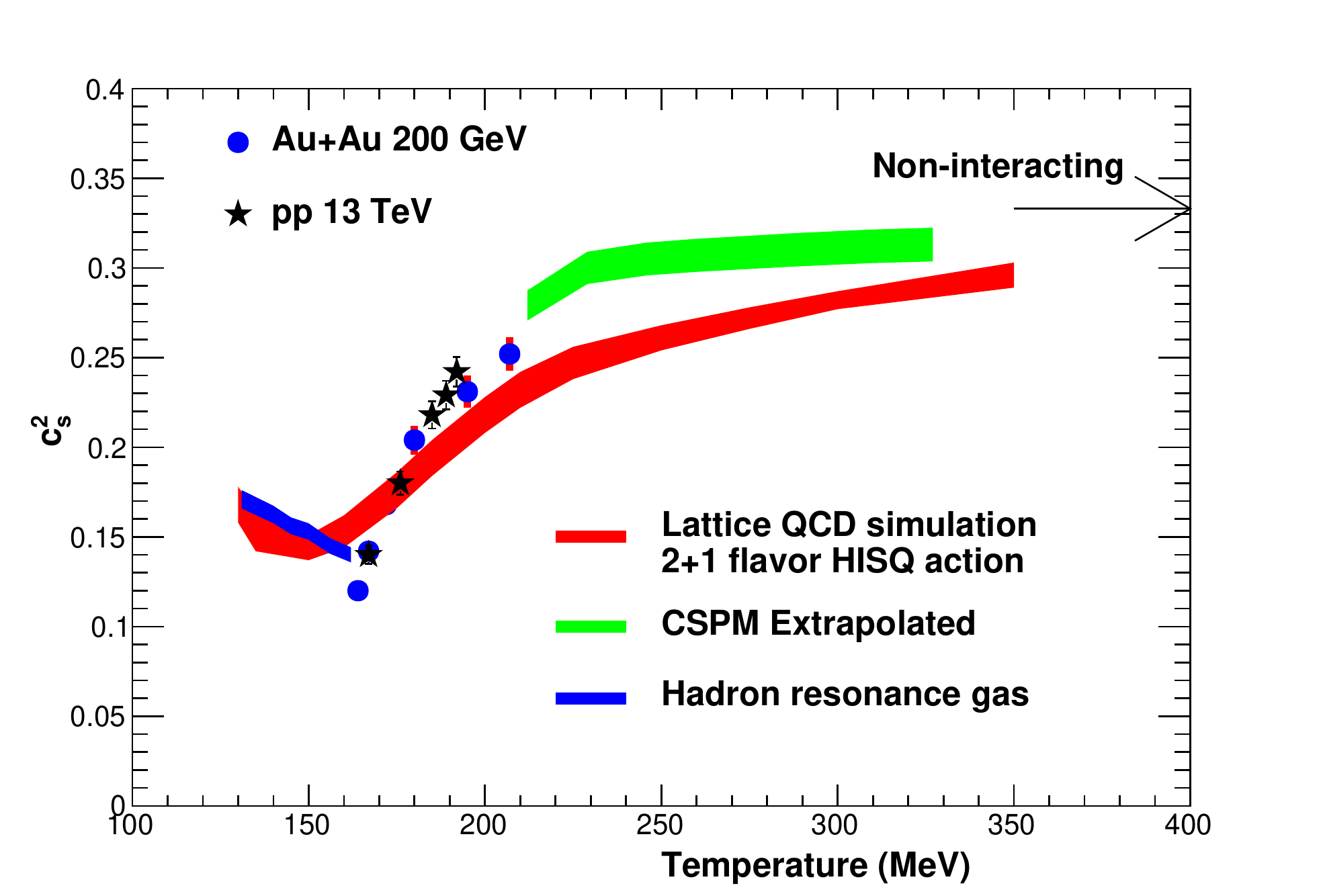}
\vspace*{0.0cm}
\caption{The speed of sound from CSPM  versus $T$ for $pp$ at 13 TeV ( black star)  and Au+Au at 200 GeV ( blue solid circle). Lattice QCD results are shown as  red band)\cite{lattice12}. Extrapolated results from CSPM ate higher temperatures are shown as green band
The physical  hadron gas with resonance mass cut off  M $\leq$ 2.5 GeV is shown as blue band \cite{eos}.} 
\label{soundfig}
\end{figure}

The QGP according to CSPM is born in local thermal equilibrium  because the initial temperature is determined at the string level. We use CSPM coupled to hydrodynamics to calculate the sound velocity. According to Bjorken boost invariant 1D hydrodynamics the sound velocity is given by \cite{bjorken} 
\begin{eqnarray}
\frac {1}{T} \frac {dT}{d\tau} = - C_{s}^{2}/\tau  \\
\frac {dT}{d\tau} = \frac {dT}{d\varepsilon} \frac {d\varepsilon}{d\tau} \\
\frac {d\varepsilon}{d\tau} = -T s/\tau ,
\end{eqnarray}
where $\varepsilon$ is the energy density, $s$ the entropy density, $\tau$ the proper time, and $C_{s}$ the sound velocity. One can eliminate $\tau$ using above expressions to obtain sound velocity as
\begin{eqnarray}
s =(1+C_{s}^{2})\frac{\varepsilon}{T}\\
\frac {dT}{d\varepsilon} s = C_{s}^{2}. 
\end{eqnarray}
Since $s = (\varepsilon + P)/T$ and $P = (\varepsilon-\Delta T^{4})/3$ one can express $C_{s}^{2}$ in terms of $\xi$
\begin{equation}
 C_{s}^{2} = \left(\frac {\xi e^{-\xi}}{1- e^{-\xi}}-1\right)\\
             \left(-\frac{1}{3} + \frac {\Delta}{12}\times \frac{1}{N} \right),
\label{soundfinal1}
\end{equation}
where $N$ is given by the dimensionless quantity $\varepsilon/T^{4}$ \cite{review}. $\varepsilon$ is the Bjorken energy density \cite{bjorken}. 
Fig.~\ref{soundfig} shows a plot of $C_{s}^{2}$ as a function of $T$ along with the lattice calculations \cite{lattice12}. The CSPM $C_{s}^{2}$ at $T \sim 200 $ MeV is 0.025, which is $\sim$ 10 $\%$ higher than the LQCD value \cite{lattice12}. The CSPM extrapolation continues to rise faster than LQCD at $ T \sim 230$ MeV, but the agreement narrows in the approach to the limiting value $T\sim$ 400 MeV. For $T\sim$ 400 MeV the CSPM and LQCD $C_{s}^{2}$ values agree within errors at $C_{s}^{2} \sim $ 0.31. 
The CSPM equation of state summarizes the data based temperature dependence of the hot nuclear matter. 

\section{Comparison with Heavy Ions}

This is the first work to obtain the initial temperature at time $\sim 1$ fm/c in $pp$ collisions at LHC energies using CSPM. 
 Since the color suppression factor  $F(\xi)$ can be normalized by the nucleon 
interaction area it is natural to compare it with the heavy ions results.
 In our earlier work $F(\xi)$ was obtained in  Au+Au collisions at 
$\sqrt {s_{NN}}$ = 200 GeV  for various centralities using STAR data \cite{review}. The results are shown in Fig.~\ref{fxipp2} along with $ {\it pp}$ (CMS) and $ {\it \bar{p}p}$ (E735)  collisions. It is observed that $F(\xi)$ as a function of  $dN_{c}/\Delta \eta$ scaled  by the transverse interaction area falls on a universal scaling curve for both hadron-hadron and nucleus-nucleus collisions. $F(\xi)$ values in high multiplicity events in $ {\it pp}$ collisions at $\sqrt s $ = 13 TeV are similar to those obtained in most central events in Au+Au collisions at $\sqrt {s_{NN}} $ = 200 GeV. This shows the importance of the string density magnitudes in these collisions.
 In Fig.~\ref{tempfig} the temperatures from Au+Au collisions at $\sqrt s_{NN} $ = 200 GeV are compared with the $pp$ collisions. In high multiplicity events for $pp$ collisions at $\sqrt s $ = 13 TeV the temperature is the same as in most central events in Au+Au.

 The $\eta/s$ values as a function of temperature from Au+Au at $\sqrt {s_{NN}} $ = 200 GeV and Pb+Pb at $\sqrt {s_{NN}} $ = 2.76 TeV collisions is shown in Fig.~\ref{viscosity}(a) along with $pp$ collisions. The $\eta/s$ value is similar both in the  high multiplicity $pp$ collisions at $\sqrt s = $ 13 TeV and in most central collisions in Au+Au at $\sqrt {s_{NN}} $ = 200 GeV. This shows that the QGP created in high multiplicity $pp$ collisions is strongly coupled. The similarity between high multiplicity $pp$ events and Au+Au indicates that the thermalization in both systems is reached through the stochastic process ( Hawking-Unruh) rather than kinetic approach (thermalization at later times)\cite{satzbook}.
\section{Summary}
We have used the Color String Percolation Model (CSPM) to explore the initial stage of high energy nucleon-nucleon collisions and determined the thermalized initial temperature of the hot nuclear matter at an initial time $\sim 1$ fm/c. In  LQCD the upper temperature limit of the non-perturbative region is $T \sim$ 400 MeV.

 In a high energy nucleon-nucleon collision many strings are produced. The high number of strings randomly overlap and the resultant color sources have higher string tension. The non-perturbative Schwinger mechanism $QED_{2}$ operates on the color sources to extract color neutral $q-\bar q$ from the vacuum. The $q-\bar q$ pairs subsequently hadronize. The net result of the coherent non-Abelian color source formation is the increase in transverse momentum and a reduction in the multiplicity.  The string formation process also produces large deceleration of the colliding nucleons and forms an event horizon. The $q-\bar q$ barrier penetration of the event horizon limits the available information and the observed pion spectra exhibit a maximum entropy temperature through the Hawking-Unruh mechanism. 

When the percolation density parameter $\xi$ = 1.2 is reached a macroscopic spanning cluster appears that describes a connected system of  $q-\bar q$ pairs. At the percolation critical transition the temperature is $T_{h}$ = 167.7 MeV. 
At lower values of $\xi$ smaller droplet formation of $q\bar q$ is indicated.
The percolation transition is known to represent a continuous phase transition. LQCD has characterized the hadron to the QGP transition as a cross-over without a latent heat. The clustering of string begins at $\sim$ 147 MeV. The pseudo-critical temperature at the center of the percolation phase transition $\sim$ 159 MeV is in reasonable agreement with the LQCD pseudo-critical value of $\sim$ 155 MeV.

CSPM is used to compute the thermodynamics of the initial stage of $pp$ collisions at LHC energies for temperature, for the shear viscosity to entropy density ratio , the trace anomaly and the sound velocity.  The data are obtained from published CMS results at $\sqrt s$ = 0.9 - 13 TeV. A universal scaling of the color reduction factor is obtained for both $pp$ and A+A. For high multiplicity events at $\sqrt {s}$ = 7 and 13 TeV  the temperature is well above the universal hadronization temperature indicating that the matter created is in the deconfined phase. 
The thermalization in both $pp$ and Au+Au is reached through the stochastic process ( Hawking-Unruh) rather than kinetic approach.
The small $\eta/s$ near the transition temperature also suggests the formation of a strongly coupled $QGP$. 

\section{Acknowledgment}
We express our thanks to N. Armesto for fruitful comments.
 C.P thanks the grant Maria de Maeztu Unit of excelence MDM-2016-0682 of Spain, the support of Xunta de Galicia under the project  ED431C 2017 and  project FPA 2017-83814 of Ministerio de Ciencia e Innovacion of Spain and FEDER. 
%
%\section*{References}

\end{document}